# Dislocation-induced anomalous softening of solid helium

Caizhi Zhou[a,b*], Jung-jungSu[a,c], Matthias J. Graf[a], Charles Reichhardt[a], Alexander V. Balatsky[a,c] and Irene J. Beyerlein[a]

[a]*Theoretical Division, Los Alamos National Laboratory, Los Alamos, New Mexico 87545, USA*

[b]*Center for Nonlinear Studies, Los Alamos National Laboratory, Los Alamos, New Mexico 87545, USA*

[c]*Center for Integrated Nanotechnologies, Los Alamos National Laboratory, Los Alamos, New Mexico 87545, USA*

*Corresponding author. E-mail: czhou@lanl.gov

**The classical motion of gliding dislocation lines in slip planes of crystalline solid helium leads to plastic deformation even at temperatures far below the Debye temperature and can affect elastic properties. In this work we propose that the gliding of dislocations and plasticity may be the origin of many observed elastic anomalies in solid $^4$He, which have been argued to be connected to supersolidity. We present a dislocation motion model that describes the stress-strain $\tau$-$\varepsilon$ curves and work-hardening rate $d\tau/d\varepsilon$ of a shear experiment performed at constant strain rate $\dot{\varepsilon}$ in solid helium. The calculated $d\tau/d\varepsilon$ exhibits strong softening with increasing temperature owing to the motion of dislocations, which mimics anomalous softening of the elastic shear modulus $\mu$. In the same temperature region the motion of dislocations causes dissipation with a prominent peak.**

Keywords: dislocation dynamics; plasticity; solid $^4$He; superfluidity

**1. Introduction**

The observation of softening of the low-temperature shear modulus in solid $^4$He with increasing temperature around 100 mK has been taken as evidence for anomalous elastic properties tied to supersolidity [1-4]. Measurements of the resonant period of torsional oscillator of solid helium showed no period drop in the case of $^3$He, while both $^3$He and $^4$He systems showed softening of the shear modulus in the same temperature range. This result suggested the importance of the role of quantum statistics and the role of defects and moving dislocations on supersolidity [5]. The connection of the motion of dislocations with supersolidity is not certain, whereas the connection of the glide of dislocations with plastic deformation of the crystal is well established [6]. The theory of strain hardening in materials science predicts the stress-strain ($\tau$-$\varepsilon$) curve and allows the study of dislocation motion and plasticity [7-8]. Such dislocation motion would strongly affect the determination of elastic properties in solid helium, a crystal with a very soft elastic shear modulus [9-10].

In this Letter, we present a new perspective on the origin of reported anomalous elastic properties of solid helium. The glide of dislocations in a dislocation forest is well understood for typical metals and can lead to large changes in the work-hardening rate (WHR) [11]. Here, we study the dynamics of classical dislocation motion and plastic behavior of solid helium. At lower temperatures one might expect that quantum dislocation motion will play a significant role [12-16]. In the case of a DC shear strain rate experiment, $\dot\varepsilon$= const., with shear stress $\tau$, we predict the softening of theWHR, $d\tau/d\varepsilon$, with increasing temperature $T$ due to dislocation glide and the creation and multiplication of dislocation loops. In the limit of $\varepsilon \to 0$, the WHR approaches the elastic shear modulus $d\tau/d\varepsilon \to \mu$ and exhibits purely elastic behavior. Our dislocation model also predicts that higher $^3$He impurity concentration in solid $^4$He pushes the onset of $d\tau/d\varepsilon$ changes to higher temperatures, because more $^3$He atoms pin more dislocations, and $^3$He pinning increases the number of immobilized dislocations, causing the crystal to harden. At the same time, the zero-temperature value of the WHR is unaffected by $^3$He atoms. Finally, the WHR decreases with increasing strain at finite temperature and attains the elastic shear modulus value at absolute zero temperature.

Our proposed scenario for dislocation dynamics in solid $^4$He differs from the prevailing view of boiling off of $^3$He impurities from dislocation lines with increasing $T$ [3], and the subsequent vibration of longer dislocation segments pinned only by the dislocation network [17-19]. In our model, dislocations glide under external loading to produce plastic strain in the material and experience interactions with the surrounding dislocation network. A main result of this work is the prediction of a dissipation peak caused by plastic deformation of the solid in the same temperature region where $d\tau/d\varepsilon$ changes most rapidly. All model predictions are remarkably similar to existing shear measurements with applied AC strain rate [1-5] and will provide a stringent test for DC strain rate experiments of solid helium.

## 2. Models and methods

We model the elastic and plastic properties of a polycrystalline sample of $^4$He in the presence of a uniform applied shear stress $\tau_{app}$ (see Fig. 1(a)). Plastic deformation arises when the dislocations in the solid start to glide. In the present model, we adopt a statistical representation of a group of dislocations, which neglects individual dislocation characteristics such as polarity, line orientation, etc. The dislocation network is then represented by a continuous distribution of dislocations and characterized by a linear density $\rho$ (total length of dislocation lines over the sample volume). The *total* dislocation density in a material consists of both *mobile* and *stored* (immobile) dislocations

$$\rho_{tot} = \rho_{mob} + \rho_{stored} \quad (1)$$

Immobile dislocations may be glissile, that is, pinned by defects, or sessile, that is, lying out of their habit plane. For example, glissile dislocations can be pinned when decorated with $^3$He atoms as sketched in Fig. 1(b). Both $\rho_{mob}$ and $\rho_{stored}$ evolve as deformation proceeds.

During dislocation glide, the increment in total dislocation density is given by [8]

$$\Delta\rho_{tot} = \Delta\varepsilon_p / bL \quad (2)$$

where $L \sim 1/\sqrt{\rho_{tot}}$ the mean-free path of the mobile dislocations and $\Delta\varepsilon_p$ is the increment of plastic strain produced by dislocation glide. The increments in stored and mobile dislocation densities are proportional to $\Delta\rho_{tot}$, as portions of an expanding dislocation loop can react with other loops and become permanently immobilized. In this study, we assume the newly added dislocations have equal chance of becoming mobile or stored. Thus, we arrive at $\Delta\rho_{mob} = \Delta\rho_{stored} = \Delta\rho_{tot}/2$.

For simplicity, we assume an average speed $v$ for all mobile dislocations in the sample volume and that their speeds are sufficiently low, such that their motion is controlled by thermal activation, obeying an Arrhenius law. In this regime, mobile dislocations glide from one pinning point to another. Their motion is resisted by $\tau_{res}$, however, a combination of thermal energy and mechanical work in excess of a resistance stress $\tau_{res}$ can overcome the Gibb's free energy of activation, $\Delta G = U_i - \tau_{eff}V$, with effective stress $\tau_{eff} = \tau_{app} - \tau_{res}$ and activation volume $V$. Accordingly, the average dislocation speed as a function of stress and temperature is given by [7-8]

$$v = bv_D \exp\left[-\Delta G / k_B T\right] \quad (3)$$

The magnitude of the Burgers vector is $b = |\mathbf{b}|$, the Debye frequency is $v_D$, the pinning energy is $U_i$, and the activation volume is $V = b^2 L$. This rate model agrees with observations in [3]. We model the resistance stress by [7-8]

$$\tau_{res} = \tau_{Peierls} + \alpha\mu b\sqrt{\rho_{tot}} \qquad (4)$$

with Peierls stress $\tau_{Peierls}$, elastic shear modulus $\mu$, and a dislocation-interaction coefficient, typically ranging from $0.1 < \alpha < 2$ [20]. In mean-field theory, the second term in (4) arises due to long-range interactions with other dislocations in the background.

The total applied strain is accommodated by both elastic and plastic strain, $\varepsilon_{tot} = \varepsilon_e + \varepsilon_p$. The plastic strain rate arising from glide obeys [6]

$$\dot{\varepsilon} = bv\rho_{mob} \qquad (5)$$

The corresponding applied shear stress $\tau_{app}$ follows Hooke's law and obeys

$$\tau_{app} \equiv \mu\varepsilon_e = \mu(\varepsilon_{app} - \varepsilon_p) \qquad (6)$$

Since the plastic work $W_p$ equals $\tau_{app}\varepsilon_p$, one should note that dislocation glide yields a finite plastic strain rate $\dot{\varepsilon}_p$ that gives rise to finite dissipation ($\dot{W}_p$).

3. Results and discussion

In this study, we assume the initial total dislocation density equals to $10^6$ m$^{-2}$ according to recent experiment and estimation [4]. The stress-strain curves in Fig. 2(a) show plastic behavior (deviation from the linear stress-strain curve) above a critical applied strain. The transition zone from elastic to plastic behavior is indicated by filled and open circles along the stress-strain curves. The applied stress effectively lowers the barrier given by the Gibb's free energy $\Delta G = U_i + \tau_{res}V - \tau_{app}V$. When $\tau_{app} < \tau_c$ only a negligible amount of the dislocations can thermally overcome the "effective" barrier. The dislocation lines are essentially pinned and the sample remains elastic. When $\tau_{app} > \tau_c$ increasing amounts of dislocations start to glide, which produces plastic strain $\varepsilon_p$ for the same applied stress (deviation from elastic behavior). Accordingly dislocation loops grow in size, thus increasing the dislocation density shown in Fig.2(b). Temperature comes into play in the elastic-plastic crossover, since our model describes dislocation gliding as a thermally activated process in Eq. (3): At higher temperature, more dislocations can overcome the barriers. This corresponds to a lower critical strain and broader transition shown in Fig. 2(a) at the higher temperature. The dislocation density then increases at the lower applied stress (strain) displayed in Fig. 2(b). To reflect the quantum nature of solid helium, the calculations here consider the case in which the pinning potential $U_i$ equals the melting temperature $T_m$ = 1.86 K [1], since we postulate that pinning originates from dislocation crossing or local melting. The other material parameters used have been reported elsewhere in the literature: $b$ = 0.364 nm, $\mu$ = 13.7 MPa [21]. Regarding the Peierls barrier, recent shear displacement experiments showed that softening of the low-temperature shear modulus happens at an applied strain larger than $2 \times 10^{-8}$ [1-3]. Since we interpret this softening instead as a consequence of dislocation glide, we use $\tau_{Peierls} = 2\times10^{-8}\mu$ in this study, which is consistent with a reported vanishing Peierls stress in Sanders' experiment [22]. Note this value is much smaller than the Peierls stress of solid $^4$He at 0K calculated via Monte Carlo simulations by Pessoa and co-workers [23]. In our model, quantum and crystal structure effects are incorporated in the effective pinning potential $U_i$ and the Peierls stress is a constant lattice resistance stress on gliding dislocations. The yielding point will change, but the key features of the predicted $d\tau/d\varepsilon$ softening and dissipation peaks will not when altering the pinning potential and the Peierls stress. Moreover, the fundamental results are not sensitive to the remaining model parameters, such as $\alpha$ =0.1.

The slope of the stress-strain curve defines the work hardening rate (WHR ≡ $d\tau_{app}/d\varepsilon_{app}$). The WHR and dissipation in Fig. 3 show a decrease in the WHR above ~100 mK accompanied by a dissipation peak. This result is strikingly similar to measurements of the dynamic shear modulus. The WHR at low temperature corresponds to that of an ideal crystal with no dissipation, since dislocation glide is negligible. At the crossover temperature, dislocations start to overcome barriers and glide. Dislocation glide contributes additional strain, which reduces the WHR while simultaneously dissipating energy, giving rise to a dissipation peak. At even higher temperatures, dislocation lines are essentially free to glide viscously; the system exhibits plastic strain but dissipates only minimal energy. We also show the WHR and dissipation for different applied strains in Fig. 3. The crossover temperature for a larger applied stress (strain) is lower since the "effective barrier" is lower in that case.

Our dislocation glide model can capture the effect of $^3$He impurities as well. Here, we assume that initially (immobile) stored dislocations are decorated by $^3$He atoms as shown in Fig. 1, and stored dislocations and $^3$He atoms are both immobile in our calculation. We model a lower $^3$He concentration by a higher initial mobile density $\rho_{mob}$ and vice versa for the same initial total dislocation density. Fig. 4 shows the WHR versus temperature. The curves for high, medium, and low $^3$He concentrations correspond to initial mobile dislocation densities of $1 \times 10^5$, $4 \times 10^5$, $9 \times 10^5$ m$^{-2}$, respectively. When the $^3$He concentration is low, more mobile dislocations are available to produce plastic deformation and the curve is sharper. When the $^3$He concentration is higher, fewer dislocations can glide to generate plastic deformation causing a broader crossover at higher temperatures. Although we neglected the "boiling off" of $^3$He atoms at high temperature, the results are still in a good agreement with the observations made in various experiments and are shown for low, medium, and high $^3$He concentrations in Fig. 4.

We next discuss the similarities and differences between our calculations for DC shear strain loading and the previous AC strain measurements of the dynamic shear modulus [1-3]. Although we model an applied DC strain rate measurement, the results share the same qualitative features with the AC experiment: both show decreases in the $d\tau/d\varepsilon$ above a crossover temperature accompanied by a dissipation peak.

The relative change in the $d\tau/d\varepsilon$ is much smaller in the AC measurement (~ 15%) than in this calculation (~ 93%). Three possible reasons may explain these differences. Firstly, our dislocation model addresses a DC strain rate ($d\varepsilon/dt$ = const) experiment, while all torsional oscillator or shear modulus experiments to date were performed under an AC strain rate condition ($d\varepsilon/dt = \omega\varepsilon_0 \cos\omega t$) or a combination thereof. These are qualitatively different experiments with very different load histories, namely constant versus cyclic loading, which is not being addressed here. Secondly, the quantitative, but not qualitative differences, between AC strain rate measurements and the DC strain rate model calculations is related to different boundary conditions. In the DC model, the shear load is homogeneously applied to the sample. Thus, the stress contributed by the externally applied load is the same for all dislocations. However, in the shear displacement experiment by Beamish's group, the transducers are encapsulated by the solid $^4$He under hydrostatic pressure. Thus, the precise stress distribution between shear transducers and the sample is very complicated. Furthermore, in this case, the measured modulus will not only be controlled by the helium sandwiched by transducers but also by contributions from the surrounding bulk helium [24]. Finally we note that grain boundaries and the distribution of multiple grain orientations can change the strength of the effect. In a polycrystalline system, the strain from different grains can act along different directions and partially cancel each other out. This would diminish the observed effect. Indeed ultrasound measurements observed a smaller change in the shear modulus of polycrystalline samples compared to single crystals. For high quality single crystals a change of 86% was reported [4]. Our model does not include such cancellation effects. This may partly explain why the AC strain rate measurement in polycrystals gives a smaller effect.

Various attempts to explain the temperature behaviour of solid $^4$He in torsional oscillator and shear modulus experiments have centered around glassy or viscoelastic models [25-28]. It is worth to mention that the viscoelastic model is a special limit of the glassy defects model and falls into the broad class of constitutive materials models that do not consider the evolution of dislocations or microstructure or creation of dislocation loops during deformation. In fact, it incorporates anelastic behavior by dash-pot elements, that means, dissipation is put in by hand in an otherwise purely elastic medium. No plastic deformation is captured by the viscoelastic model. On the contrary, our dislocation density model is a mechanical based model that directly evolves a dislocation density and proliferation of dislocation loops. In this view, the viscoelastic model would not be suitable for describing dislocation induced plastic deformation of solids.

## 4. Conclusions

In summary, we proposed a simplified isotropic density-based dislocation model to describe qualitatively many essential features of the anomalous low-temperature shear modulus of solid $^4$He measured with an AC strain rate. More importantly, we predicted many new stress-strain curves, work hardening rates, and dissipation properties for a DC strain rate shear experiment. In our model, the classical glide of dislocations in slip planes under load has no relationship with supersolidity.

One may speculate that the existence and dominance of dislocation induced plastic deformation facilitates a nonuniform supersolid state, because a supersolid state is possible in $^4$He along the core of static dislocations [29,30]. An interesting aspect of a highly tangled dislocation network may be its local loss of crystalline order, owing to strong lattice distortions by closely arranged dislocation cores [31]. Those distorted regions are more amenable for superfluidity and are a potential candidate for a nonuniform supersolid phase. The dynamic effects of plastic deformations and dislocation glide discussed here will play a role in the dynamic measurements of solid $^4$He regardless of the presence or absence of superfluidity. In the absence of more detailed microstructural information the model proposed here offers an effective method to check whether the dislocation motion is the fundamental mechanism for plasticity in solid helium at the lowest temperatures.

**Acknowledgements**

We acknowledge discussions with J. Beamish and C. J. Olson-Reichhardt. This work was supported by the U.S. DOE at Los Alamos National Laboratory under contract No. DE-AC52-06NA25396 and the Office of Basic Energy Sciences (BES). CZ acknowledges support provided by the Center for Nonlinear Studies, Statistical Physics Beyond Equilibrium Project from the Los Alamos National Laboratory Directed Research and Development Office.

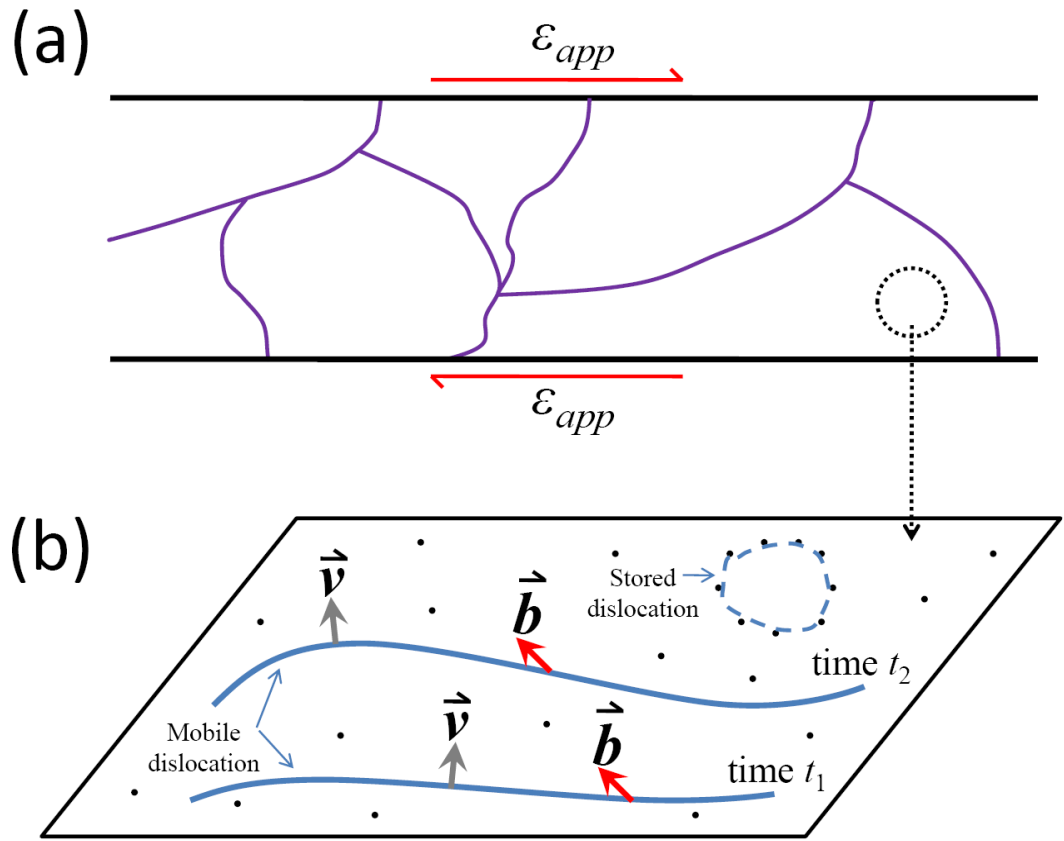

**Fig. 1:** (Color online). (a) Illustration of the applied shear stress on the solid $^4$He sample with curved lines indicating grain boundaries; (b) magnification of the circled region in (a) and plot of a mobile dislocation moving on the slip plane under external load. Black dots indicate $^3$He pinning points, solid lines are mobile dislocations, and dashed lines are stored (immobile) dislocations.

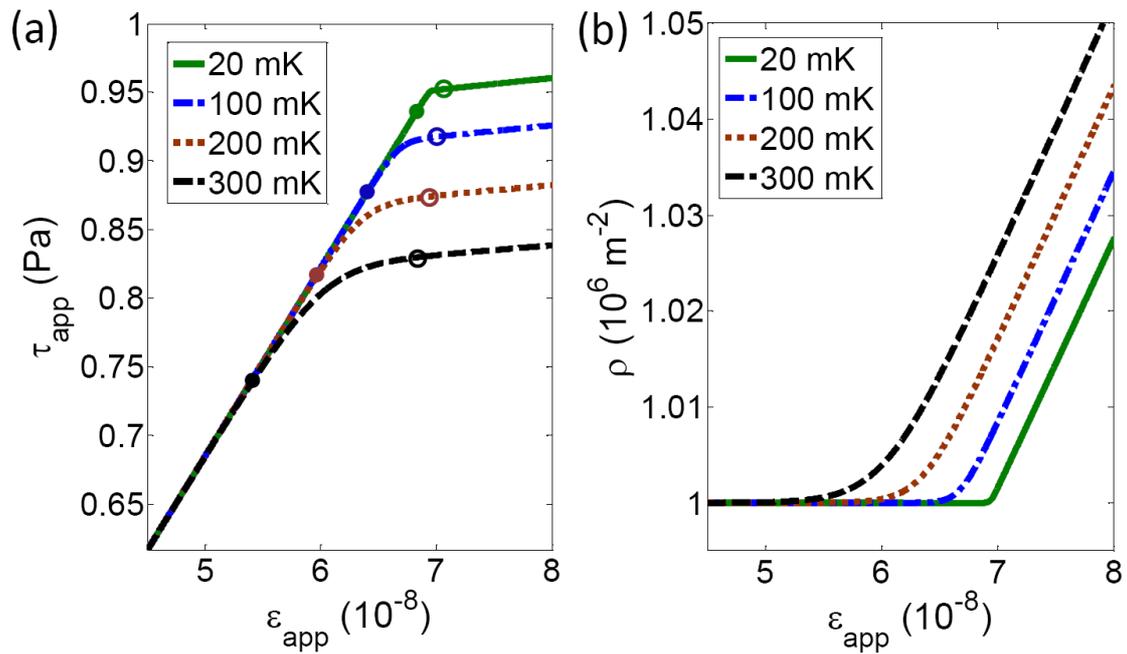

**Fig. 2:** (Color online). (a) Stress-strain curves for different temperatures at an applied strain rate of $\dot{\varepsilon} = 0.001$ s$^{-1}$ with initial total and mobile dislocation densities $10^6$ m$^{-2}$ and $10^5$ m$^{-2}$, respectively. The filled and open circles bracket the transition zone between elastic and plastic behavior. (b) Plot of the corresponding dislocation density-strain curves of (a). As temperature decreases, the transition point is delayed to higher shear strain and the transition zone narrows and sharpens.

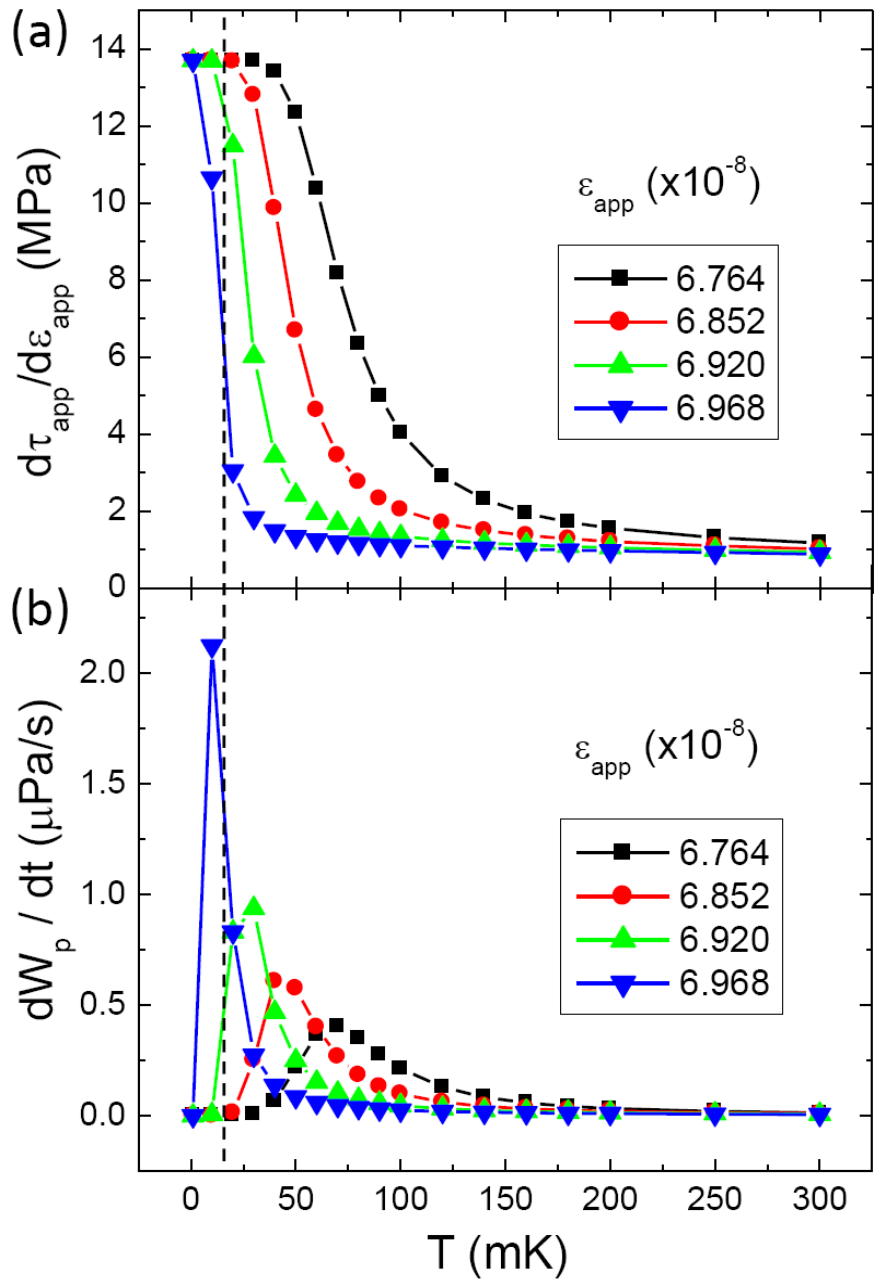

**Fig. 3**: (Color online). (a) Work hardening rate versus temperature for different strains. The parameters are the same as in Fig. 2; (b) corresponding plastic dissipation of (a). The dashed line at $T = 20$ mK marks the cut-off of a typical low temperature experiment. For thermally activated dislocation glide the work hardening rate approaches the elastic shear modulus at $T = 0$ K and no plastic work is dissipated.

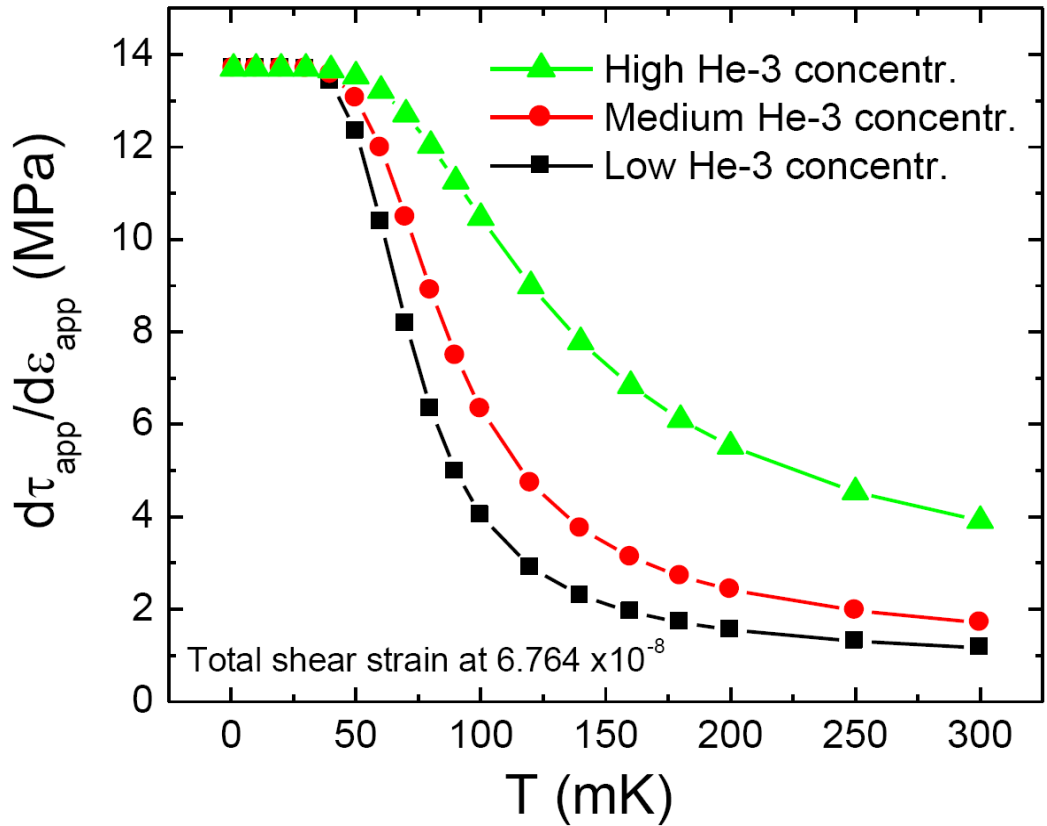

**Fig. 4**: (Color online). Work hardening rate versus temperature for different pinning point ($^3$He) concentrations. The curves for high, medium, low $^3$He concentrations correspond to initial mobile dislocation densities of $1\times10^5$, $4\times10^5$, $9\times10^5$ m$^{-2}$, respectively. All calculations have the same initial total dislocation density of $10^6$ m$^{-2}$ and total strain amplitude of $\varepsilon = 6.764 \times 10^{-8}$.